\input phyzzx
\Frontpage
\centerline{ARE THERE TOPOLOGICAL BLACK HOLE SOLITONS IN STRING THEORY?}
\foot{e-mail address: mazur@psc.psc.sc.edu}
\vfill
\centerline{Pawel O. Mazur}\foot{present address: Department of Physics and 
Astronomy, University of South Carolina, Columbia, SC 29208, USA.}
\vskip .2in
\centerline{Syracuse University, Physics Department,}
\centerline{Syracuse, N.Y. 13244}
\vskip .5cm
\centerline{February 1987}
\vskip .5cm
\centerline{published in Gen. Rel. Grav. 19 (1987) 1173.}
\foot{Gravity Research Foundation Essay for 1987.}
\vfill
\centerline{Abstract}
\vskip .2in
We point out that the celebrated Hawking effect of quantum instability of
black holes seems to be related to a nonperturbative effect in
string theory. Studying quantum dynamics of strings in the gravitational
background of black holes we find classical instability due to emission of
massless string excitations. The topology of a black hole seems to play
a fundamental role in developing the string theory classical instability
due to the effect of sigma model instantons. We argue that string theory
allows for a qualitative description of black holes with very small masses
and it predicts topological solitons with quantized spectrum of masses.
These solitons would not decay into string massless excitations but could be
pair created and may annihilate also. Semiclassical mass quantization
of topological solitons in string theory is based on the argument showing
existence of nontrivial zeros of beta function of the renormalization group.
\vfill
\endpage
\doublespace
   It is believed that black holes should play
a very important role in quantum gravity. The most remarkable theoretical
prediction about black holes, which also opened the beautiful connection with
thermodynamics is the Hawking effect.$^{1,3}$ Hawking discovered that a black
hole will radiate quantum mechanically with a thermal spectrum having a
temperature related to the hole's mass $M$, angular momentum $L$, and the
electric and magnetic charges $Q$ and $P$.
\par
   Classical general relativity and the physically reasonable assumption of
nonnegativity of energy imply that the event horizon of a black hole has the
topology of a sphere $S^2$. It is plausible to assume that black holes
``survive'' quantization of gravity and they should be present also
in the future theory of all interactions. At least it is so in the (super)
string theory.$^4$ A black hole is the highly nontrivial and nonperturbative
solution of the Einstein equations which is classically stable and
unique.$^{2}$ It is impossible to obtain a black hole solution by summing up
a finite number of terms in the expansion of Einstein's equations around
Minkowski vacuum, a fact which is demonstrated by the fact that the
topology of spacetime outside the event horizon, or, what is the same, the
topology
of the Euclidean black hole is $R^2\times S^2$. In the following we will
define a black hole in semiclassical or quantum gravity by saying that the
second homology group of its manifold is nontrivial, $H_2({\it M}_{bh})=Z$.
\par
    What are the implications of this beautiful mathematical fact in quantum
gravity? It seems that the physical implications of this simple fact are not
quite well understood. The condition of the nontrivial second homology group
$H_2({\it M}_{bh})$ for black holes is reminiscent of the nonabelian
t'Hooft-Polyakov
magnetic monopoles which are present if the second homotopy group
of a certain manifold is nontrivial. In particular it seems unclear if there is
some relation between the fundamental quantum mechanical instability of black
holes and the fact that $H_2({\it M}_{bh})=Z$. One may ask, is it possible to
stabilize a black hole with respect to the Hawking effect by introducing
topological charges which would bound the energy of a hole from below
(the Bogomolny'i bound)? If it were possible, then there would exist
topological black hole solitons. The usual way to understand the Hawking
effect in QFT does
not rely on the fact that the topology of an event horizon is $S^2$. Rather, it
is important that there be a horizon and that the Bekenstein-Hawking entropy
be $S_{bh}={A\over 4}$, where $A$ is the area of the event horizon. It seems
that this
property of black hole physics, which attributes an intrinsic entropy to a
black hole and equals it to a basic geometrical quantity---the area $A$ of the
generator of the second homology group of a black hole manifold---should be
derived from fundamental principles of quantum gravity.
Accepting the information theory
point of view on entropy, advocated particularly by Bekenstein$^6$, we reach
the conclusion that the number of ways a black hole can be ``built up'' depends
on its geometrical properties only: $\Gamma=e^{S_{bh}}=e^{A/4}$.
\par
   I think that there should exist the way to understand the relation between
the topological properties of black holes, like $H_2({\it M}_{bh})=Z$, their
entropy and
fundamental quantum mechanical instability. The Hawking effect should manifest
its presence in the future unified theory of gravity and all fundamental
interactions.
\par
   One of the promising candidates for such a theory is theory of strings.$^4$
What is attractive in this theory is the presence of a massless spin-2 particle
in its perturbative spectrum. The tree level interactions of this ``graviton''
are
effectively described by the Einstein-Hilbert action. In the classical limit
of large occupation number of gravitons we have a nontrivial macroscopic
graviton ``condensate'' $\Phi_{\mu \nu}$, and we can define the effective
metric $g_{\mu \nu}=\eta_{\mu \nu}+\Phi_{\mu \nu}$ which is a solution of the
Einstein equations. The conformal invariance implied by the geometrical fact
that the
string world-sheet is two-dimensional is highly restrictive in QFT.$^4$ In the
low energy limit only the massless string excitations play a role and we can
study the dynamics of the quantum string propagating in the classical background
of the massless field ``Bose condensates''. It is definitely interesting to
address
the question of how the quantum dynamics of strings is restrictive for black
holes. The closed string contains other massless excitations: a spin-0 dilaton
and a spin-0 abelian two-form $B_{\mu \nu}$, which plays a fundamental role in
anomaly cancellations.
One may expect that the Hawking effect will manifest itself in quantum
mechanical inconsistency of string propagation in the gravitational field of a
black hole. We would like to demonstrate in this essay how the nontrivial
topology
of black holes implies their instability in string theory. The same sort of
argument leads to the conclusion that there are possible topological black hole
solitons stabilized by the topological ``quantum numbers'' and they would be
solutions of string equations of motion. They would be created or annihilated
in
pairs. Also, what seems to be the fundamentally new aspect of black hole physics
brought about by string theory, the hole's mass or irreducible mass
$\left(A\over 16\pi \right)^{1/2}$ is
quantized. This sort of behavior of Planck's mass black holes was postulated
several years ago by Bekenstein.$^{3,6}$ The famous Bekenstein-Hawking entropy
formula obtains for such objects with the simplicity of combinatorics.
\par
   Let us now go to the details and see how string theory makes use of
the nontrivial
black hole topology. The string interaction with background fields is
described by the nonlinear sigma model
$$I = {1\over 4\pi \alpha '}\int d^2\sigma \sqrt{h}
(h^{\alpha \beta}\partial_{\alpha}X^\mu \partial_{\beta}X^\nu g_{\mu \nu}(X)
+\epsilon^{\alpha \beta}\partial_\alpha X^\mu \partial_\beta X^\nu
B_{\mu \nu}(X)+{1\over 2}R\Phi(X)),\eqno(1)$$
where $h_{\alpha \beta}$ is the string world sheet metric and
$(2\pi\alpha')^{-1}$=$M_S^2$ is the string tension.
Considering only the gravitational background and demanding that the
two-dimensional nonlinear sigma model (1) be quantum mechanically conformally
invariant leads, in the lowest approximation, to the classical Einstein
equations for the background metric.
\par
   Classical string equations of motion are derived from the condition of
conformal invariance of quantum dynamics of the sigma model---the sigma model
should be Weyl invariant on the quantum level!
Black holes are solutions of the string equations of motion in the perturbative
expansion in $\alpha'$. Is this also true nonperturbatively?
The answer is no, because black holes are unstable due to the Hawking effect;
we conclude that instability of black holes in string theory is a genuinely
nonperturbative effect in $\alpha'$! Let us see how it occurs.
String theory gives us the unique opportunity of including topological issues
of quantum gravity into the game. Breakdown of Weyl invariance or beta function
depends only on the short-distance behavior of the QFT and does not depend
on the string world-sheet topology. The Euclidean path integral
$Z=\int DX Dh\ exp(-I)$ can be calculated on $S^2$, when the Weyl invariance
holds, which corresponds
to the classical approximation to string theory. The condition for a classical
solution in string theory is $\langle V\rangle=0$, where $V$ is the vertex
operator of a given
physical state. It means that the vacuum is stable with respect to emission of
particles corresponding to $V$. In order to study stability of a given ground
state it is sufficient to check it at the tree level, which corresponds to the
classical approximation of string theory. The last condition corresponds at the
tree level to the world-sheet conformal invariance. Conformal invariance on
$S^2$
means, in particular, invariance under the scaling $z\mapsto\lambda z$ on the
complex plane.
A physical closed-string vertex operator has dimension two and transforms as
$V\mapsto \lambda^{-2}V$, which implies $\langle V\rangle=0$. $V$ should be
evaluated only at zero momentum,
which is on shell only for massless particles. If $\langle V\rangle\not=0$
for
massless particles, it means that a given classical ground state is unstable
with respect
to emission of ``soft'' particles into the vacuum. If it happens for black
holes
it means that the Hawking effect is a semiclassical phenomenon in string theory.
\par
    Consider now a black hole background on which a quantum string propagates
which has a
characteristic length scale $R=\left(A\over4\pi\right)^{1/2}$; $A$ is the
area of the event horizon. It is convenient to rescale
the sigma model fields $X^\mu\mapsto RX^\mu$  introducing dimensionless
fields and the action
$$I = {1\over 2g^2}\int d^2\sigma\ \partial_\alpha X^\mu\partial^\alpha X^\nu
g_{\mu \nu}(X),\eqno(2)$$
where $g^2={\pi\alpha'\over 2R^2}$ is the dimensionless sigma model coupling
constant.
The condition for con\-formal invari\-ance of (2) is the van\-ish\-ing
of the beta func\-tion:
$\beta(g^2)=-(\mu{\partial \over\partial \mu})g^2(\mu)$. One may ask if the
Schwarzschild black hole is a solution of the string equations of motion or,
equivalently, if it is stable.
Of course we assume that the $d=10$ dimensional string model is compactified
on some
compact manifold $K$ such that its beta function is vanishing, which seems to
restrict the possible compactifications ${\it M}_{bh}\times K$. The sigma model
(2) on a
black hole background has nontrivial instantons due to the fact that
$H_2({\it M}_{bh})=\pi_2({\it M}_{bh})=Z$ is nontrivial and there exist maps
$X^\mu(\sigma)$ such that $I$
is finite and bounded from below. A historic role of instantons is to violate
classically valid symmetry or produce effects which couldn't be seen at any
finite order of perturbative expansion. So one may ask, what is the effect of
world sheet instantons on a black hole background?
\par
   The effect we find is destabilization of a black hole. We can evaluate the
partition function for the sigma model on the Schwarzschild black hole
background
in the instanton (WKB) approximation. In this approximation we can also tell
what the beta function is. The Euclidean rescaled Schwarzschild metric in
Kruskal coordinates is
$$ds^2 = g_{\mu\nu} dX^{\mu}dX^{\nu}=f{du}d{\bar u}+r^2(1+w{\bar w})^{-2}
dwd{\bar w},\eqno(3)$$
$w=cot(\theta/2)e^{i\phi}$, $u=(r-1)^{1/2}e^{(r+2ikt)/2}$,
$f=r^{-1}e^{-r}$, and $k$ is the surface gravity $k={1 \over4M}$.
\par
    Instantons are complex (holomorphic or antiholomorphic) maps from $S^2$
to the $u$ or $w$-plane. Only $u=const$ instantons contibute to the partition
function $Z$,
because they only have finite action. Small oscillations around the $w$-sector
$CP(1)$
instantons lead to the classical Coulomb gas partition function, a standard
result of the $CP(1)$ sigma model. Small oscillations around $u=const$ give
rise to a
contribution depending on $r=const$ and the integral in the collective variable
$r$ has the effect that only $r=1$ contributes significantly to the path
integral
($r=1$ is a saddle point!). Basically, we would expect that the main
contribution
to the string partition function would come from strings ``living'' close to
the event horizon. A similar phenomenon was discovered by Thorne and Zurek in
their statistical derivation of $S_{bh}={A\over 4}$.
Now we employ the fact that the dynamics of strings in a black
hole background is governed by the $CP(1)$ sigma model on the event horizon
(here is where the topology of a black hole enters) and observe that it
does have a nonvanishing beta function. The $CP(1)$ sigma model is
asymptotically free,
which means that $g=0$ is an ultraviolet fixed point corresponding to
$R\mapsto\infty$ or $M_{bh}\mapsto\infty$. It means that only ultraheavy
black holes are stable in
classical string theory. For very large black hole masses the Hawking effect
can be neglected or, equivalently, a black hole metric is a solution of the
classical string equations of motion. This dual point of view reflects the fact
that the Hawking effect seems to correspond in string theory to the
instability of a black hole due to massless particle emission.
\par
   Is it possible to produce an example of vanishing beta function for nonzero
$g$? In QFT it would correspond to the absence of the Hawking effect. It is
only possible when the Bekenstein-Hawking temperature vanishes. Quantum
mechanically stable black holes do exist and they are
the extremal charged Kerr black holes. For simplicity we consider only the
Reissner-Nordstr\"om case: $M_{bh}=M_{Pl}(Q^2+P^2)^{1/2}$. The physical
interpretation of the
vanishing temperature is quite simple because we have a state, stabilized by the
absolute charge conservation, behaving like a soliton. However,there is
a problem with interpreting the quantum mechanically stable black holes as
solitons. The mass of a soliton is semiclassically quantized. We would like to
argue that string theory offers a natural mechanism for semiclassical mass
quantization of black holes.
\par
   Let me give here first an ad hoc but instructive demonstration of how the
quantization of mass of the extremal Reissner-Nordstr\"om dyons can occur.
Assume that the theory has magnetic monopoles, which are in fact dyons (this
happens in string theory due to the topological effects of the $B$ field$^4$)
satisfying the Dirac-like quantization condition: $QP=n/c$, $c$ an integer.$^4$
Then a black hole mass is: $M_{bh} = M_{Pl} (Q^2 +{n^2\over Q^2c^2})^{1/2}$.
However, the ground state of a soliton should have a minimal energy;
minimization of $M_{bh}$ gives: $Q^2 = n/c$, $Q=P$, $M_{bh}^2 = 2M_{Pl}^2n/c$. 
We observe that {\it the classical formula for a black hole dyon mass is 
duality invariant} but quantum effects need not respect this classical 
symmetry. The irreducible Christodoulou mass of a semiclassically 
quantized R-N dyon and its Bekenstein-Hawking entropy are:
$M_{ir} =( A/16\pi)^{1\over2}=M_{Pl}\left(n\over2c\right)^{1\over2}$,
$S_{bh} = 2\pi  n/c$.
The number of internal states which
give rise to this entropy is:$^6$ $\Gamma = e^{S_{bh}} = e^{2\pi n\over c}$.
Now we may ask
in how many possible ways we can produce the mass level with the topological
``quantum number'' $n$ ?  A simple combinatorics gives $\Gamma = 2^{n-1}$,
from which the information theory entropy is $S=ln\Gamma=(n-1)ln2$. Discarding
an additive constant these two entropies naively agree if ${2\pi\over c}=ln2$,
or $c={2\pi\over ln2}=9.03$.
For this value of $c$ the lowest lying soliton state would
have a mass  $M_{bh}=.5 M_{Pl}$. This quite crude argument suggests that there
may exist topological black hole solitons with masses of the order of $M_{Pl}$
which
are stable quantum mechanically with respect to emission of massless particles
not carrying the same charges as solitons. Such topological solitons may be
created and annihilated in pairs with opposite charges, but these processes can
be only studied in full quantum gravity. It is amusing to observe that the
Dirac-like quantization condition for fractional charges arises in string
theory where $c$ is simply, as Witten observed, the order of the fundamental
group of $K$.
The topological properties of a compact manifold $K$ have implications for low
energy physics. In particular the Euler characteristic in some models is
related to the
number of generations of fundamental fermions.$^4$ The nontrivial first
homotopy
group of $K$ reduces the Euler characteristic $c$ times, where $c$ is the rank
of $\pi_1(K)$. Is it only a coincidence that the mass spectrum of black hole
solitons
is related to the topological properties of a compact manifold $K$ which occurs
in string theory compactifications?
\par
It is important to recognize that string theory offers the possibility of
stable topological black hole solitons, which by definition are objects
satisfying string equations of motion---the sigma model beta function
vanishes for these configurations!
\par
How do we get a vanishing beta function in QFT? It is a very rare situation
when the beta function can be calculated exactly and has isolated zeroes.
The topological effects of the $B$ gauge field may lead to an effectively
free field theory in two dimensions. Such an effect is known and a beautiful
example
was discovered by Witten$^5$---the nonlinear sigma model on a group manifold
with the Wess-Zumino term. For certain values of the coupling constant
$g^2={4\pi\over n}$ the sigma model is equivalent to a theory of free fermions
and the beta function has
zeroes at these values of $g$; $n$ is the quantized coupling constant of the
topological WZ term. In fact, there is a nontrivial Wess-Zumino term on the
black hole background. Its existence is strictly related to the fact that the
horizon of Euclidean black holes has $S^2\times S^1=Y$ topology. There is
a WZ term if
the third homology group of $Y$ is nontrivial: $H_3(Y)=Z$. It means also that
there
is a closed but not exact form on $Y$ which locally can be written as $dB$.
An
example is easily constructed: $dB=dt\wedge \omega$, where $t$ is an ``angle''
on $S^1$ and $\omega$ is
a volume 2-form on $S^2$. $B$ is then given locally as $t\omega$. This leads
to quantization of the coefficient of the WZ term. By standard arguments the
topological coupling is unrenormalized. It means that when the ``metric''
coupling
$g$ is renormalized it depends on the ``bare'' coupling $g_0$ and the integer
WZ
coupling constant $n$. It may happen then that the beta function will have
isolated
zeros which asymptotically are $g^2\sim1/n$. This would lead to the
quantization
condition for the characteristic length scale of a black hole: $R^2\sim n$.
But now we have two mass scales in the problem: $M_{Pl}$ and $M_S$.
\par
    Concluding, we would like to point out that the Hawking effect seems
to be related to a semiclassical effect in string theory
but is completely nonperturbative in $\alpha'$
due to instantons. The  $R^2\times S^2$ topology of black holes plays a
fundamental role
in establishing their instability in string theory. String theory allows for
a qualitative description of black holes with small masses and seems to
predict topological black hole solitons which would not decay into string
theory massless ``mesons''. One may expect that those objects have a size
comparable to the compactification scale and therefore would be effectively
10-dimensional.
\par
This research was supported by NSF grants PHY 83-18350 and PHY 86-12424
with Relativity Group at Syracuse University.

{\it Note added on December 18th, 1996.}

It makes no sense for me to list here the numerous papers which have appeared 
in the last decade on the subject of strings and black holes, 
``axionic black holes'', ``quantum hair'', ``world-sheet instantons'', 
exact solutions in the multidimensional context, etc..  
Instead, I would like to suggest that 
the basic assumptions about the subject of so-called black hole entropy 
in the context of strings be thought over again. 
Does Nature really require strings and in what limit? 

\endpage
\singlespace
\centerline{REFERENCES}
\vskip .2in
\item{1)} Hawking, S.W., (1975), Commun. Math. Phys. {\bf43}, 199.
\item{2)} Mazur, P.O., (1982), J. Phys. {\bf A15}, 3173.
\item{3)} Bekenstein, J.D., (1974), Lettere al Nuovo Cimento {\bf 11}, 467.
\item{4)} Green, M.B., Schwarz, J.H. and Witten, E., (1987)
   {\sl Superstring Theory}, Vols. I and II
   (Cambridge University Press, Cambridge).
\item{5)} Witten, E., (1984), Commun. Math. Phys. {\bf92}, 455.
\item{6)} Bekenstein, J.D., (1981), In {\sl To Fulfill a Vision: Jerusalem
   Einstein
   Centennial Symposium on Gauge Theories and Unification of Physical
   Forces}, (Addison-Wesley Publishing Co., New York).
\end